\documentclass[Journal]{copernicus}

\usepackage{url,lineno}
\usepackage{color}
\usepackage[T1]{fontenc}

\begin{document}

\title{Fractional Laplace Transforms - A Perspective
}

\author[1,2]{R. A. Treumann\thanks{Visiting the International Space Science Institute, Bern, Switzerland}}
\author[3]{W. Baumjohann}

\affil[1]{Department of Geophysics and Environmental Sciences, Munich University, Munich, Germany}
\affil[2]{International Space Science Institute, Bern, Switzerland}
\affil[3]{Space Research Institute, Austrian Academy of Sciences, Graz, Austria}

\runningtitle{$\kappa$-Laplace Transform}

\runningauthor{R. A. Treumann and W. Baumjohann}

\correspondence{R. A.Treumann\\ (rudolf.treumann@geophysik.uni-muenchen.de)}

\received{ }
\revised{ }
\accepted{ }
\published{ }


\firstpage{1}

\maketitle

\paragraph*{\bf{Abstract}}
A form of the Laplace transform is reviewed as a paradigm for an entire class of fractional functional transforms. Various of its properties are discussed. Such transformations should be useful in application to differential/integral equations or problems in non-extensive statistical mechanics.


\section*{Introduction}
The importance of functional transforms as a tool in all fields of theoretical physics cannot be overestimated. As for an example, the two complementary representations of quantum physics are related via the Hilbert space equivalent of Fourier transformations. Generally speaking, functional transforms switch from the original space in which a function is given to its complementary (or dual) space, where the function, after having been transformed, can possibly be treated in a simpler way.   

Functional transformations have been introduced by Fourier and Gauss. Fourier used them to expand any given function into an infinite sum of trigonometric functions showing that their superposition reproduced the original function to an arbitrary degree of precision. His method led to the wide field of spectral representations and turbulence theory. Gauss applied the concept of functional transformation to represent measurements of the gravitational and geomagnetic fields on Earth's surface into sums of Legendre polynomials and spherical harmonics. He also demonstrated how the expansions could be used to solve the Laplace and Poisson equations thereby paving the way for the application of functional transforms to the solution of the differential and integral equations of mathematical physics including boundary value problems. This concept also lies at the bottom of perturbation theory. Laplace, in application of Fourier's transformation to initial value problems, realized that temporary problems were unidirectional and thus required substantial modification. 

Meanwhile, several other kinds of transformations have been discovered \citep[cf., e.g.,][]{abram}. In recent years, in particular in relation to so-called non-extensive statistical mechanics, a whole industry developed in reformulating a large class of exponential or logarithmic function based mathematical functions into related functions which in some asymptotic limit approach either exponentials or logarithms. This development is interesting as, by simple analogue, it creates new functions available for application \citep[for reviews the reader is directed to][and references therein]{diaz2005,mansour2009,mansour2012,liva2013}. 

To do justice to history we mention that the $q$-transformation on which non-extensive statistical mechanics is based as well as several of the modified $q$-functions were suggested a century ago \citep{jackson1904a,jackson1904b,jackson1904c,jackson1905}. Elaborated versions of all those related to the hypergeometric function and other polynomials can be found already in \citep{hahn1949}.  There, a version of $q$-modified Laplace transforms is even given in passing. The focus on these functions is purely mathematical though some of them have found application in non-extensive statistical mechanics \citep[cf., for a listing,][]{liva2013}. We also note that some very interesting mathematical investigations of the $q$-Laplace transforms have recently been undertaken \citep{chung2013a,chung2013b} uncovering some hitherto unknown connections to the operator calculus.   

The present note directs attention to a different version of modifying the Laplace transform in view of enlarging the available mathematical tools. Our formulation uses $\kappa$ representations instead of the above mentioned  so-called $q$-deformed functions \citep[cf., e.g.,][]{lenzi1999}. In fact, the transformation between both representations is simply that $q=1+1/\kappa$ \citep{treumann1997,liva2013}. Since various possibilities exist for obtaining exponentials as asymptotic limits, there are several ways of defining transformations. `$q$-deformed' transforms exploit raising the $q$-deformed exponential to some power $s$ identified as the Laplace variable. Such $q$-Laplace transforms differ in several respects from the one given below. Which one should be better suitable, depends on the problem under consideration. In non-extensive statistical mechanics the $q$-deformed Laplace transform seems appropriately tailored. We note that a similar construction is possible also for Fourier transforms where, however, complications arise due to the explicit complexity of the modification.  

\subsubsection*{Definition.}
Introducing the real free parameter $\kappa\in\textsf{R}$, we define the fractional or $\kappa$-Laplace transform as
\begin{equation}
{L}_\kappa[f(t);r]\equiv g_\kappa(s,r)=\int\limits_0^\infty\frac{\mathrm{d}t\ f(t)}{\left(1+st/\kappa\right)^{\kappa+r}}
\end{equation}
with inverse transform
\begin{equation}
{L}_\kappa^{-1}[g(s,r)]\equiv f(t)=\frac{1}{2\pi i}\int\limits_{c-i\infty}^{c+i\infty}\frac{\mathrm{d}s\ g_\kappa(s)}{\left(1+st/\kappa\right)^{-\kappa-r}}
\end{equation}
and convention that the index on the transform ${L}_\kappa$ refers to $\kappa$ in the exponent. These forms are obtained by replacing the exponentials in the ordinary Laplace transform and its inverse by the generalized Lorentzians $\left(1\pm st/\kappa\right)^{-\kappa -r}$. The philosophy in mind is that for \emph{any rational} numbers $\kappa, r\in \textsf{R}$ the generalized Lorentzians, for $\kappa\to\infty$ and $r$ fixed, asymptotically approach exponentials $\exp(\mp st)$. One verifies this readily by applying l'Hospital's rule. 
The Lorentzians in the transforms are functionally inverse while not being inverse functions. In fact the inverse functions would be obtained inverting them with respect to either $s$ or $t$ which, however, would cause loss of the property of a functional transform. It would not  reproduce the Laplace transforms at $\kappa\to\infty$, contrasting other uses of the Lorentzian replacement found in the literature, in particular Tsallis' non-extensive statistical mechanics \citep[for its fairly complete review cf., e.g.,][]{tsallis2004}.

The above transformations are inverse to each other satisfying the relation $L_\kappa^{-1}L_\kappa=\mathsf{I}$, with $\mathsf{I}$ the identity operator. This is verified performing the operator product yielding another representation of the $\delta$ function 
\begin{equation}
\delta(t-t')=\frac{1}{2\pi i}\int\limits_{c-i\infty}^{c+i\infty}\!\!\! \mathrm{d}s\ \mathrm{e}^{(\kappa+r)\log\left[1+{s\left(t'-t\right)}/{\left(\kappa+st\right)}\right]}
\end{equation}
For $\kappa\to\infty$ it reproduces the normal Laplace form of the Dirac function, thus yielding the representation of $\delta(t)$ appropriate for our purposes.

\subsubsection*{Properties.}
That the fractional operator ${L}_\kappa[\alpha U(t)+\beta V(t)]$ is linear need not to be proven. In order to demonstrate that the above transformations are indeed functional transformations we apply the fractional Laplace transform to the function $f(t)=\delta_a(t)=1/a$ with $a=$ const and check its behavior at $\lim_{a\to0}$. Inserting for $f(t)$ into $L_\kappa$ yields
\begin{eqnarray}
{L}_\kappa[\delta_a(t);r]&=&\frac{1}{a}\int_0^a\frac{\mathrm{d}t}{\left(1+st/\kappa\right)^{\kappa+r}} \nonumber\\
&&\\
&=& \frac{a}{a}F\left(\kappa+r,1;2;-\frac{sa}{\kappa}\right)\nonumber
\end{eqnarray}
where $F(\alpha,\beta;\gamma;z)$ is the hypergeometric function, which for $z=0$ becomes $F(\alpha,\beta;\gamma;0)=1$. Hence, $\lim_{a\to0} g_\kappa(\delta_a)=1$, and the fractional Laplace transform indeed trivially reproduces the property of the Dirac-$\delta$ function $\delta(t)$ for all $\kappa, r$. One thus obtains trivially $L_\kappa[\delta(t)]=1$. 
Transforming instead the Heaviside function $H(t)$ yields  that ${L}_\kappa[H(t);r]=\kappa/[s(\kappa+r)]$ which for $r=0$ simplifies to the obvious result ${L}_\kappa[H(t);r]=1/s$ since $\delta(t)$ is its derivative.
Moreover, the transform of $f(at)$ becomes simply
\begin{equation}
{L}_\kappa[f(at);r]= a^{-1} g_\kappa(s/a,r)
\end{equation}

Not all properties of the Laplace transform survive when making the transition to its fractional sister. For instance adding a constant $c$ to $s$ does not lead to any useful result. This is unfortunate as it indicates that the fractional Laplace transform is not restrictionless applicable to  functions of period $T$. Indeed, when taking the transform of a periodic function $f(t)=f(t-T)$ with $f_0(t)=f(t)$ for $0\leq t<T$ and $f_0(t)=0$ elsewhere, i.e. $f(t)=f_0(t)+f(t-T)$ for $t\geq0$, one obtains instead that
\begin{eqnarray}
\tilde{{L}}_{\kappa}[f(t);r]&\equiv&\left\{{L}_\kappa- \left(\frac{\kappa}{\kappa+sT}\right)^{\kappa+r}{L}_{\kappa-sT}\right\}[f(t);r]\nonumber\\
&&\\
&=&{L}_\kappa[f_0(t);r]\nonumber
\end{eqnarray}
where 
\begin{eqnarray}
{L}_{\kappa-sT}[f(t);r]&=&\int\limits_0^\infty\frac{\mathrm{d}t\ f(t)}{\left(1+st/\kappa\right)^{\kappa+r-sT}}, \\ 
{L}_{\kappa}[f_0(t)]&=&\int\limits_0^T\frac{\mathrm{d}t\ f_0(t)}{\left(1+st/\kappa\right)^{\kappa+r}}
\end{eqnarray}
Since ${L}_{\kappa-sT}\neq{L}_\kappa$, these relations cannot be simplified anymore. 

\subsubsection*{Derivatives.}
When applying the fractional Laplace transform to the derivative $f'(t)$ of the function $f(t)$ one obtains
\begin{eqnarray}
{L}_\kappa[f'(t);r]&=&\int\limits_0^\infty\frac{\mathrm{d}t\ f'(t)}{\left(1+st/\kappa\right)^{\kappa+r}} \nonumber\\
&& \\
&=&-f(0)+(\kappa+r)\frac{s}{\kappa}\int\limits_0^\infty\frac{\mathrm{d}t\ f(t)}{\left(1+st/\kappa\right)^{\kappa+r+1}}\nonumber
\end{eqnarray}
Hence
\begin{equation}
{L}_\kappa[f'(t);r]=-f(0)+s\left(1+\frac{r}{\kappa}\right){L}_{\kappa+1}[f(t);r]
\end{equation}
The fractional Laplace transform applied to the derivative of a function increases its order by one unit. Application to the second derivative $f''(t)$ then becomes
\begin{eqnarray}
{L}_\kappa[f''(t);r]&=&-f'(0) -s\left(1+\frac{r}{\kappa}\right)f(0)+\nonumber\\
&&\\
&&+s^2\left(1+\frac{r}{\kappa}\right)\left(1+\frac{r+1}{\kappa}\right){L}_{\kappa+2}[f(t);r]\nonumber
\end{eqnarray}
Further iteration yields for the fractional Laplace transform of the $n$th derivative $f^{(n)}(t)$ the expression
\begin{eqnarray}
{L}_\kappa[f^{(n)}(t);r]&=&-\sum\limits_{\ell=0}^{n-1}\frac{\Gamma(\kappa+r+\ell)}{\Gamma(\kappa+r)}s^\ell f^{(\ell)}(0)+\nonumber\\
&&\\
&&+\frac{\Gamma(\kappa+r+n-1)}{\Gamma(\kappa+r)}s^n{L}_{\kappa+n}[f(t);r]\nonumber
\end{eqnarray}
where
\begin{equation}\label{eq-kappan}
{L}_{\kappa+n}[f(t);r]=\int\limits_0^\infty\frac{\mathrm{d}t\ f(t)}{\left(1+st/\kappa\right)^{\kappa+r+n}}\equiv L_\kappa[f(t);r+n]
\end{equation}
The increasing power with $n$ in the denominator of the integrand assures that the fractional Laplace transforms of higher order derivatives exist if only the derivative of $f(t)$ exists to this order. 

Since addition of $n$ can be understood simply as replacement $r\to r+n$, the Laplace transform of derivatives behaves pseudo-recursive. This pseudo-recursive behavior distinguishes it from the ordinary Laplace transform. Application to differential equations will thus not lead to simple linear sets of equations for the Laplace transformed function but  produces a set of pseudo-recursive relations which are distinguished by $r$-dependent factors as is easily shown for a polynomial $f(t)$. 

\subsubsection*{Convolution theorem.}
When dealing with products of functions of the form $f(\tau)g(t-\tau)$ where $f(\tau)=0$ for $\tau<0$ and $g(t-\tau)=0$ for $t<0$ then the ordinary Laplace transform implies that the Laplace transform of the product becomes the product of the Laplace transforms of the functions $f$ and $g$. Applying the fractional Laplace transform to the convolution product under the same conditions yields now
\begin{equation}
{L}_\kappa\left[\int \mathrm{d}\tau f(\tau)g(t-\tau);r\right]= \kappa {L}_{\kappa-1}[g(t);r]\ast{L}_\kappa[f(t);r]
\end{equation}
where ${L}_{\kappa-1}$ is of the form given in Eq. (\ref{eq-kappan}). This expression suggests that the transform of a convolution indeed remains a product of two Laplace transforms. However, the order of the transform of the convoluted function $g(t-\tau)$ has become reduced by one unit. In addition the fractional transform is multiplied by $\kappa$.  The right-hand side of the last formula can also be rewritten by attributing the change in the index to the constant $r$:
\begin{equation}
L_\kappa \left[f\ast g\right]=\kappa L_\kappa \left[ g; r-1\right]\ast L_\kappa \left[ f;r\right]
\end{equation}
Conversely the last formula defines the product of two Laplace transforms as the transform of a folding integral. For this product to exist warranting convergence, the free parameter $r$ must be chosen sufficiently large. 

\subsubsection*{Remarks.}
The $\kappa$-Laplace transform proposed in this note is just one form of modified Laplace transformations. So far, regarding their mathematical properties \citep{chung2013a,chung2013b} and application \citep[for transforms of various functions see, e.g.,][]{lenzi1999}, the literature makes use of the $q$-modified versions of Laplace transforms, first proposed long ago by Hahn \citep{hahn1949}. These applications are particularly suited for problems in non-extensive statistical mechanics \citep{tsallis2004}. 

The $\kappa$-Laplace transform given here, though being different, is a relative of the $q$-Laplace transform. Because of its simpler form it might be easier to handle. Whether it is appropriate or not depends on the problem to which it can be applied. It will not be difficult to simply calculate transformed versions of a number of known functions and tabulate them, as was done for the $q$-transform \citep{lenzi1999}. More interesting is, however, the identification of the class of problems to which the $\kappa$-transform can be applied.  

The most interesting property of the $\kappa$-Laplace transform is its pseudo-recursive character when applied to derivatives of functions. This makes them valuable in application to differential equations including fractional differential equations \citep[cf., e.g.,][]{pod,rah}, though the ordinary Laplace transform provides already a useful tool for their treatment, in the case of fractional differential equations just generating some additional factors. This is completely satisfactory already and well known from the fractional versions of the diffusion equation, known as Fick's equation \citep[cf., e.g.,][]{sokolov2002}, which leads to so-called subdiffusion with mean-square displacements increasing slower than linear with time. Applying the $\kappa$-Laplace transform would generate another additional factor. Whether this provides any advantage, is questionable. It will depend strongly on the disposed problem. Super-diffusion based on L\'evy flight dynamics, is the opposite to sub-diffusion. It might be a candidate for application of the $\kappa$-Laplace transform to the underlying Fokker-Planck equation in phase space. So far attempts to describe super-diffusion are based only on turbulent probability spectra of the form $p(\mathbf{k})\propto \exp(-k^\alpha)$ with $\alpha<2$ \citep[see, e.g.,][]{tsallis2004}. Application of such spectra has also been given to the collisionless magnetic reconnection problem which is important in space physics \citep{treumann2014}
. 

In any case, the apparent simplicity of the $\kappa$-Laplace transform advertises it for being used in physical problems. Genuinely mathematical investigations of the kind given in \citep{chung2013a,chung2013b} are as well very welcome as they enlighten the internal structure of the transformation and its relation to operator calculus.

\begin{acknowledgements}
This research was part of a Visiting Scientist Program at ISSI, Bern in 2006/2007. Hospitality of the librarians Andrea Fischer and Irmela Schweizer, and the technical administrator Saliba F. Saliba,  is acknowledged. RT thanks the referees for clarifying comments and suggestions of related literature.
\end{acknowledgements}











\begin{thebibliography}{99}
\bibitem[Abramowitz \& Stegun(1972)]{abram} Abramowitz M \& Stegun I A (1972) Handbook of Mathematical Functions, Dover Publ., Inc., New York, Ch. 29

\bibitem[Chung(2013a)]{chung2013a}Chung W S (2013) On the q-Laplace transform in the  non-extensive statistical physics, arXiv:1301.5480v1 [physics.gen-ph]

\bibitem[Chung(2013b)]{chung2013b}Chung W S \& Kim T (2013) On the q-analogue of Laplace transform, arXiv:1307.6752v1 [math.NT], and references therein

\bibitem[Diaz \& Teruel(2005)]{diaz2005} Diaz R \& Teruel C (2005) q,k-generalized gamma and beta functions, J Nonlin Math Phys 12, 118-134, doi:10.2991/jnmp.2005.12.1.10

\bibitem[Gell-Mann \& Tsallis(2004)]{tsallis2004} 
Gell-Mann M \& Tsallis C, eds. (2004) Nonextensive Entropy - Interdisciplinary Applications, Oxford University Press, Oxford UK  

\bibitem[Hahn(1949)]{hahn1949} Hahn W (1949) Beitr\"age zur Theorie der Heineschen Reihen, Die 24 Integrale der hypergeometrischen q-Differenzengleichung. Das q-Analogon der Laplace-Transformation, Math. Nachr. 2, 340-379

\bibitem[Jackson(1904a)]{jackson1904a} Jackson F H (1904a) The application of basic numbers to Bessel's and Legendre's functions, Proc London math Soc II 2, 192-220

\bibitem[Jackson(1904b)]{jackson1904b} Jackson F H (1904b) A basic-sine and cosine with symbolical solution of certain differential equations, Proc Edinburgh math Soc 22, 28-39

\bibitem[Jackson(1904c)]{jackson1904c} Jackson F H (1904c) Pseudo-periodic functions analogous to the circular functions, Messenger Math  34, 32- 39

\bibitem[Jackson(1905)]{jackson1905} Jackson F H (1905a) The application of basic numbers to Bessel's and Legendre's functions, Proc London math Soc II, 3, 1- 20

\bibitem[Lenzi et al.(1999)]{lenzi1999} Lenzi E K, Borges E P \& Mendes R S (1999) A q-generalization of Laplace transforms, J Phys A: Math Gen 32, 8551-8561, doi:10.1088/0305-4470/32/48/314 

\bibitem[Livadiotis \& McComas(2013)]{liva2013}
Livadiotis G \& McComas D J (2013) Understanding kappa distributions: A toolbox for space science and astrophysics,  Space Sci Rev {175}, 183-214, doi:10.1007/s11214-013-9982-9

\bibitem[Mansour(2009)]{mansour2009} Mansour T (2009) Determining the k-generalized gamma function $\Gamma_k(x)$ by functional equations, Int J Contemp Math Sci 4, 1037-1042, doi:10.1007/s11005-008-0290-3 

\bibitem[Mansour \& Shabani(2012)]{mansour2012} Mansour T \& Shabani A S (2012) Generalization of some inequalities for the ($q_1, \dots, q_s$)-gamma function, Mathematiche (Catania) 67, 119-130

\bibitem[Podlubny(1999)]{pod} Podlubny I (1999) Fractional Differential Equations, Academic Press, San Diego-London, Ch. 4


\bibitem[Rahimy(2010)]{rah} Rahimy M (2010) Applications of fractional equations, App Math Sci 4, 2453

\bibitem[Sokolov et al.(2002)]{sokolov2002} Sokolov I M, Klafter J \& Blumen A (2002) Fractional kinetics, Phys Today (November issue) 48-54, doi:10.1063/1.1535007

\bibitem[Treumann(1997)]{treumann1997} Treumann R A (1997) Theory of superdiffusion for the magnetopause, Geophys Res Lett 24, 1727-1730, doi:10.1029/97GL01760

\bibitem[Treumann \& Baumjohann(2014)]{treumann2014} Treumann R A \& Baumjohann W (2014) Superdiffusion revisited in view of collisionless reconnection, Ann Geophys 32, in press


\end{thebibliography}
\end{document}